# Semantic and Task-Oriented V2X Communications: Pushing the Limits of V2X Networks Scalability

Luca Lusvarghi*, Javier Gozalvez, Mohammad Irfan Khan, Seyhan Ucar, Miguel Sepulcre, and Onur Altintas

*Abstract*–**Scalable Vehicle-to-Everything (V2X) networks are key to support the large-scale deployment of connected and automated mobility. However, the scalability of V2X networks is currently challenged by the limitations of existing V2X communication paradigms, which prioritize the reliable and timely delivery of the transmitted information over a careful message content selection – an approach that can potentially lead to the transmission of unnecessary information and an inefficient usage of communication resources. Semantic and task-oriented V2X communications have recently been proposed to address these scalability challenges by focusing on the content of the transmitted messages, particularly on its relevance to the intended receivers. In this paper, we numerically demonstrate that semantic and task-oriented V2X communications can substantially improve the scalability of V2X networks, increasing by up to a 4.1x factor the number of supported vehicles under high-density conditions. In addition, we show that semantic and task-oriented V2X communications can also decrease the inter-reception time between consecutive messages by up to 67% and lead to a twofold increase in the probability of successfully delivering all required relevant information to the intended receivers.**

## 1. INTRODUCTION

Connected and Automated Vehicles (CAVs) will leverage Vehicle-to-Everything (V2X) communications to share basic information such as their position, heading, and speed, as well as richer information such as their planned trajectories and locally detected objects (e.g., other vehicles or vulnerable road users), cooperatively extending their situational awareness beyond their onboard sensors' field of view. An enhanced V2X-enabled situational awareness is key to improving the vehicles' understanding of the driving environment and, ultimately, maximize their driving safety and efficiency. Currently, CAVs exchange information according to V2X communication paradigms primarily designed to prioritize the reliable and timely delivery of the transmitted information. This design principle underpins the majority of existing communication systems and corresponds to a Level A (*technical level*) design according to Shannon and Weaver's three-level communication model [1]. However, a Level A design pays limited attention to the content of the transmitted messages and may therefore lead to the transmission of information that is not needed by the receiving vehicles. The transmission of unnecessary information can lead to an inefficient usage of the available communication resources that can compromise the scalability of V2X networks. Scalable V2X networks are fundamental to support the large-scale deployment of V2X-enabled connected and automated driving services on limited spectrum resources while guaranteeing an enhanced situational awareness at the communicating vehicles. To overcome the scalability challenges of existing Level A communication paradigms, substantial research efforts have recently been devoted in the context of 6G to the design of novel Level B (*semantic level*) and Level C (*effectiveness level*) communication paradigms focused on the content of the transmitted messages.

Level B semantic communications concentrate on the meaning of the transmitted information. Meaning refers to the essential set of features that needs to be transmitted to allow a correct interpretation of the data at the receiver. Exemplary applications of the semantic communication paradigm were presented in [2]-[4], focusing on the vehicle-to-vehicle exchange of images in unicast communication settings. In [2], the authors proposed a semantic communication system

This work was supported by the European Union under the 2023 MSCA Postdoctoral Fellowship program (project no. 101153845). This work was also partially funded by MICIU/AEI/10.13039/501100011033, "ERFD/EU" (PID2023-150308OB-I00), and Generalitat Valenciana (CIAICO/2024/167).

Luca Lusvarghi, Javier Gozalvez, and Miguel Sepulcre are with the Networked Systems Lab, Universidad Miguel Hernandez de Elche, Elche, 03202, Spain. E-mail: {llusvarghi, j.gozalvez, msepulcre}@umh.es. Mohammad Irfan Khan, Seyhan Ucar, and Onur Altintas are with Toyota Motor North America R&D, InfoTech Labs, Mountain View, CA, 94043, USA. E-mail: {mohammad.irfan.khan, seyhan.ucar, onur.altintas}@toyota.com.

*Corresponding author.

that can achieve higher image compression ratios than conventional lossy and lossless compression techniques while improving image reconstruction quality at the receiver. In [3] and [4], channel coding and image compression were focused on the most meaningful regions of the input image (e.g., those containing vehicles or other objects). These works show that, in unicast settings, semantic communications can greatly decrease the amount of transmitted information by focusing on the transmission of meaning rather than raw data. However, V2X communications are inherently broadcast and the design of semantic V2X communication systems able to transmit the specific meaning required by multiple intended receivers is a challenging research problem that has not been fully explored yet.

Level C task-oriented communications focus on identifying and transmitting only the information required by the receiver to execute its tasks. In the V2X domain, task-oriented communications have been mainly proposed in the context of cooperative perception[1], where CAVs exchange information about locally detected objects (e.g., their position, speed, and heading) to collaboratively overcome their onboard sensors line-of-sight and range limitations. These task-oriented solutions rely on the Value of Information (VoI) of detected objects to discriminate between unnecessary (low VoI) and necessary (high VoI) information for the receivers. For example, in [5], CAVs curate the content of their messages leveraging a VoI definition based on the dynamics of detected objects, prioritizing the transmission of objects with larger position, speed, or heading variations (i.e., higher VoI). The contributions in [6], [7], and [8] rely on VoI definitions based on the objects' entropy, detection uncertainty, and detection accuracy, respectively. Additional VoI definitions based on the object's redundancy or classification confidence have also been included in [9]. VoI-based task-oriented communications can decrease the number of objects included in each message with respect to baseline approaches where vehicles transmit all locally detected objects [10]. Yet, they are not able to completely prevent the transmission of unnecessary information, as existing VoI definitions only partially consider the intended receivers' context. Context refers to the driving and communication environment in which a piece of information is exchanged or processed. It is defined by, for example, the receiver's current and future state (position, heading, and speed), previously exchanged V2X messages, and the mobility of surrounding objects. Context is fundamental to estimate if a piece of information will impact the intended receiver's task and, therefore, if it is worth being transmitted. For example, information about a high speed (high dynamics-based VoI [5]) and non-redundant (high redundancy-based VoI [9]) object crossing an intersection might not be needed by a receiving vehicle that is leaving the intersection (context), as it has no impact on its driving.

A novel semantic and task-oriented V2X communication paradigm for broadcast V2X communications has been recently proposed by the authors in [11]. The proposed paradigm features a joint Level B and Level C design and is centered on the relevance of the exchanged information: transmitting vehicles curate the content of their messages based on the relevance of the transmitted information to their intended receivers. The relevance of a piece of information captures the context-dependent impact that its meaning (Level B) has on the driving of the intended receivers (Level C). By accounting for the intended receivers' context, relevance more accurately captures the impact of the transmitted information on the intended receivers' tasks and allows a more accurate identification of unnecessary and necessary information. This paper advances the state-of-the-art by assessing the potential scalability gains of semantic and task-oriented V2X communications in conventional broadcast scenarios with a single or multiple V2X channels. In the single-channel scenario, vehicles exchange different messages and types of information (their position and speed, their detected objects, and their planned trajectories) on a single V2X channel. In the multi-channel scenario, there is a dedicated V2X channel for the exchange of each type of message or information. The single-channel scenario is the initial V2X deployment option considered in both the EU and the US, while the multi-channel scenario is currently under consideration for future deployment phases [12]. The obtained results numerically demonstrate that semantic and task-oriented V2X communications can push V2X network scalability well beyond its current limits. In particular, simulation results show that semantic and task-oriented V2X communications can increase by up to a 4.1x factor the number of supported vehicles in high-density single-channel scenarios, and that these scalability gains are maintained in multi-channel scenarios. We show that these scalability gains stem from the improved communication efficiency of semantic and task-oriented V2X communications, which can nearly double the probability to successfully serve a vehicle with the required relevant information. Furthermore, we analyze the sensitivity of the scalability gains to different driving conditions and show that, compared to existing communication approaches, semantic and task-oriented V2X communications can decrease the inter-reception time between messages by up to 67%, ensuring a more regular and frequent exchange of information that is key for the support of V2X-enabled safety applications.

[1] Cooperative perception is the terminology employed in ETSI (European Telecommunications Standards Institute). In SAE (Society of Automotive Engineers) terminology, cooperative perception is known as sensor data sharing.

The rest of this paper is organized as follows. Section 2 outlines the operation of semantic and task-oriented V2X communications and qualitatively illustrates their benefits through a cooperative perception example, while Section 3 details the scenario and system model employed in this study. Section 4 presents the scalability analysis results obtained in the single-channel and multi-channel scenarios, and discusses the main challenges towards the practical deployment of semantic and task-oriented V2X communications. Lastly, Section 5 draws the conclusions.

## 2. Semantic and Task-Oriented V2X Communications

In the relevance theory [13] developed by linguists D. Wilson and D. Sperber, the search for relevance is highlighted as the key communication principle that allows human interactions to achieve the greatest possible communication efficiency. The search for relevance refers to the natural human ability to identify – and communicate – only the most relevant information required to convey a desired message, thereby minimizing the communication efforts. The relevance of a piece of information is defined as its impact on the listener's perception of the world when processed within a context of existing assumptions [13]. Semantic and task-oriented V2X communications make the search for relevance the core of vehicular communications, bringing its highly efficient relevance-based communication principles to the V2X domain. In semantic and task-oriented V2X communications, the relevance of a piece of information is defined as the context-dependent impact that its meaning has on the receiving vehicle's understanding of the driving environment and, ultimately, on its driving [11].

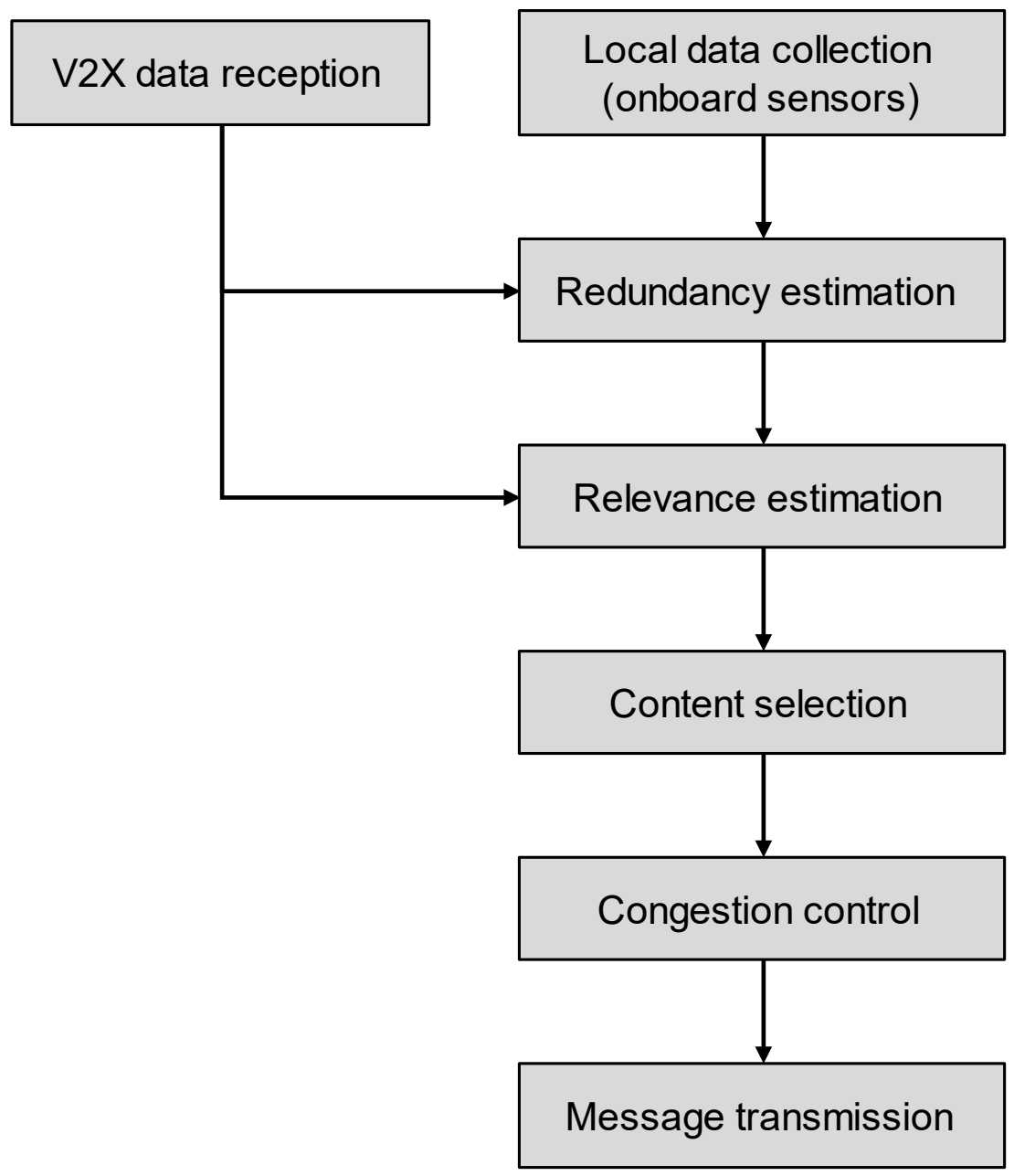

**Fig. 1**. Semantic and task-oriented V2X communications.

### *2.1. Communication Principles*

In semantic and task-oriented V2X communications, CAVs curate the content of the transmitted messages based on its relevance to their intended receivers. The operation of a transmitting vehicle is illustrated by the block diagram in Fig. 1. A transmitting vehicle first estimates which locally collected information is redundant for its intended receivers (*redundancy estimation* in Fig. 1). To do so, it checks the content of messages received over the V2X network, assuming that the content of any correctly decoded message is already available, i.e., redundant, at all intended receivers. This redundancy estimation approach is commonly used in cooperative perception [9]. Redundant information is assumed to not improve the intended receivers' understanding of the driving environment and, therefore, to have no impact on their driving. For this reason, information estimated to be redundant is considered irrelevant and is filtered out from the transmitted messages. Then, the transmitting vehicle processes the available contextual information to estimate the context-dependent relevance of non-redundant locally collected information for all its intended receivers (*relevance estimation* in Fig. 1). Relevance estimation consists of identifying which information can have an impact on the driving of each intended receiver, based on its context. The implementation of relevance estimation techniques has been preliminarily discussed in [11] and is discussed in greater detail in Section 4.4. Contextual information refers to any data, collected by the transmitting vehicle through its onboard sensors or via V2X, that can be leveraged to estimate the intended receivers' context. In the V2X domain, contextual information includes:

- Information about the road layout, speed limits, and driving rules. In general, any information describing the driving environment.
- The current position, speed, and heading of the intended receiver. More generally, any information about the current and future state of the intended receiver, including its planned trajectory.
- Previously exchanged messages containing information about the intended receiver's surrounding environment or any information about the data already available at the intended receiver.
- Information about the V2X communication link, including the channel load and the transmitter-receiver distance.

In broadcast scenarios, the relevance of each piece of locally collected information is estimated considering all intended receivers. The multi-receiver relevance estimate obtained for each piece of locally collected information is leveraged to select the content of the transmitted messages (*content selection* in Fig. 1). A transmitting vehicle includes in its message only the information that is relevant to at least one of its intended receivers. Prior to their transmission, messages are always checked by the congestion control mechanism (*congestion control* in Fig. 1). Congestion control might drop a message, i.e., prevent its transmission, to ensure that the network load remains within pre-defined limits.

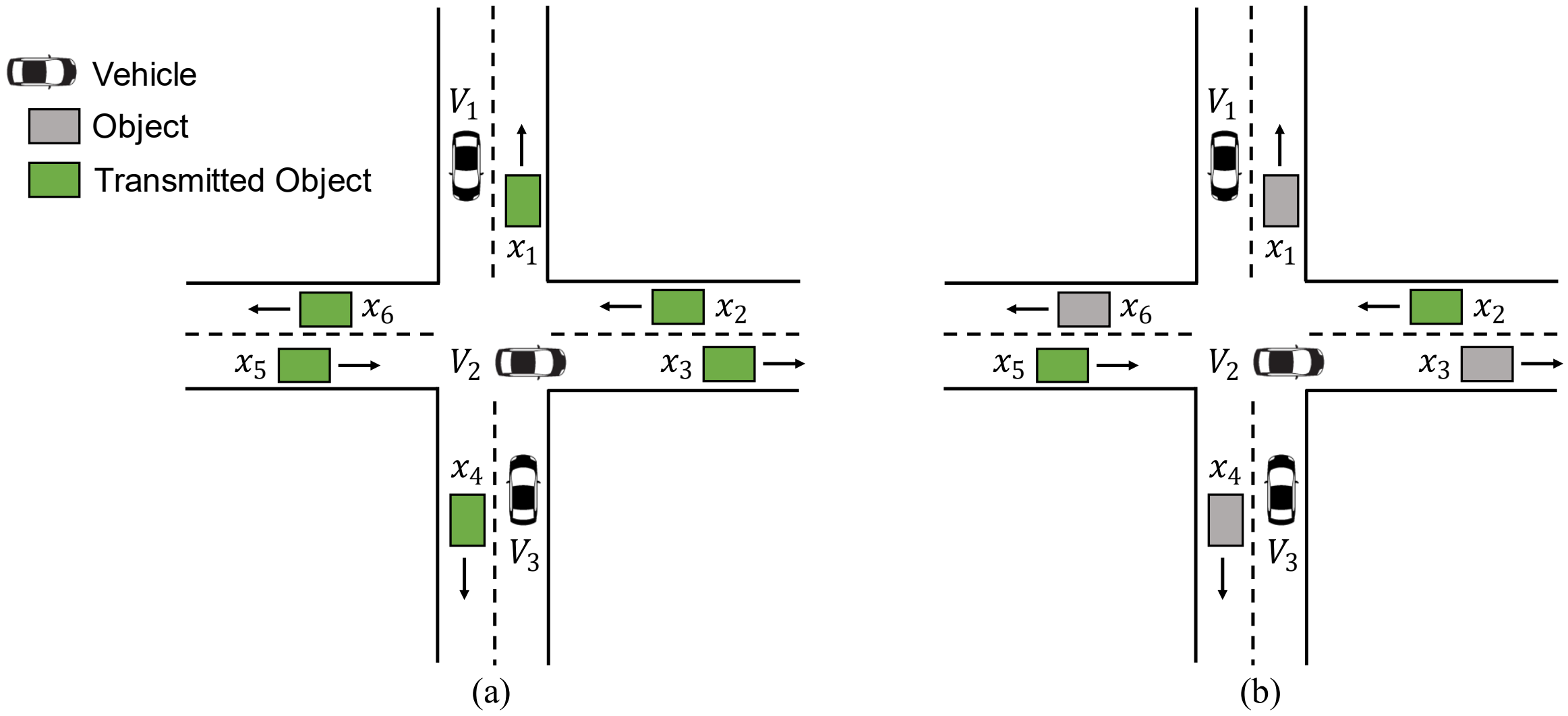

**Fig. 2**. Cooperative perception without (a) and with (b) semantic and task-oriented V2X communications.

## *2.2. Cooperative Perception Example*

We illustrate the operation and benefits of semantic and task-oriented V2X communications considering the cooperative perception example reported in Fig. 2. Fig. 2 represents a 4-way intersection with three vehicles ($V_1$, $V_2$, and $V_3$) and six objects ($x_1$ to $x_6$). In this example, we assume that $V_1$ and $V_3$ are the receiving vehicles and that $V_2$ is the transmitter. We further assume that $V_2$ can detect all six objects with its own onboard sensors and that the objects represent non-connected vehicles.

In Fig. 2(a), $V_2$ selects the content of its transmitted message according to the baseline approach supported by both ETSI [9] and SAE [14] cooperative perception standards. The baseline approach allows vehicles to include all locally detected objects in their transmitted messages. As a result, in Fig. 2(a), $V_2$ includes all six detected objects in its transmitted message. Note that Fig. 2(a) could also represent the operation of the VoI-based redundancy mitigation approach defined by ETSI in [9]. With redundancy mitigation, detected objects deemed to be redundant (i.e., already reported by another vehicle) are assigned a low VoI and excluded from the transmitted messages. In Fig. 2(a), $V_2$ would still transmit all six detected objects as long as they have not already been reported by any other vehicle.

In Fig. 2(b), $V_2$ employs the semantic and task-oriented V2X communication approach and, therefore, selects the content of its transmitted message based on the relevance of each detected object to its intended receivers ($V_1$ and $V_3$). In this case, only objects $x_2$ and $x_5$ are included in the transmitted message. Objects $x_2$ and $x_5$ are relevant to both $V_1$ and $V_3$ because they are approaching the intersection and might influence the driving of $V_1$ and $V_3$, which are also approaching the intersection. The remaining objects ($x_1$, $x_3$, $x_4$, and $x_6$) are all leaving the intersection and cannot have any impact on the driving decisions of either $V_1$ or $V_3$. For this reason, they are deemed irrelevant to both $V_1$ and $V_3$ by $V_2$ and, therefore, excluded from the transmitted message.

The comparison between Fig. 2(a) and Fig. 2(b) qualitatively demonstrates that semantic and task-oriented V2X communications can greatly improve the communication efficiency with respect to existing cooperative perception approaches by reducing the amount of transmitted information (i.e., objects) while still guaranteeing the transmission of the most relevant information required to enhance the intended receivers' situational awareness. The comparison also suggests that VoI-based communication approaches centered on the objects' redundancy may not fully capture the driving needs of the intended receivers, potentially leading to the transmission of unnecessary (or irrelevant) information.

TABLE I

SUMMARY OF EMPLOYED NOTATION AND ACRONYMS.

| **Symbol/Acronym** | **Description** |
|---|---|
| $N$ | Total number of connected vehicles |
| $K$ | Total number of exogenous variables |
| $V_n$ $(n = 1, \dots, N)$ | Connected vehicle |
| $x_k$ $(k = 1, \dots, K)$ | Exogenous variable |
| $P(D_{k,n})$ | Detection probability |
| $D_{k,n}$ | Distance between $x_k$ and $V_n$ |
| $D_{max}$ | Relevance range. Maximum distance between relevant ex. vars. and $V_n$ |
| $R$ | Total number of relevant ex. vars. within $D_{max}$ |
| CAM | Cooperative Awareness Message |
| MCM | Maneuver Coordination Message |
| CPM | Cooperative Perception Message |

## 3. SCENARIO AND SYSTEM MODEL

We analyse the scalability of V2X networks considering a driving scenario populated by $N$ connected vehicles and $K$ exogenous variables. Exogenous variables represent typical driving environment objects (e.g., other non-connected vehicles, trucks, pedestrians, etc.) and are randomly distributed according to a 2D Poisson Point Process (PPP). The probability that a vehicle $V_n$ $(n = 1, \dots, N)$ detects an exogenous variable $x_k$ $(k = 1, \dots, K)$ is modelled by a detection probability $P(D_{k,n})$ that depends on $D_{k,n}$, the distance between $V_n$ and $x_k$, and captures typical onboard sensors' detection capabilities.

At each vehicle $V_n$, we assume that an exogenous variable can be either relevant or irrelevant. A relevant exogenous variable represents an object (e.g., another vehicle, an obstacle, a vulnerable road user) that affects the driving of $V_n$, while an irrelevant exogenous variable represents an object that does not have any impact on $V_n$'s driving decisions. To capture realistic driving conditions, we limit the maximum distance between $V_n$ and a relevant exogenous variable by a relevance range $D_{max}$. All exogenous variables located beyond $D_{max}$ are irrelevant for $V_n$. Within $D_{max}$, the total number of relevant exogenous variables is set equal to $R$ and their density around $V_n$ is a linearly decreasing function of the distance from $V_n$. This design choice reflects typical real-world driving conditions, where relevant objects that may influence a vehicle's driving decisions are more likely to be located in its proximity.

Each vehicle $V_n$ generates three different types of messages. The first message type represents the Cooperative Awareness Messages (CAMs) or Basic Safety Messages (BSMs) respectively defined in ETSI [15] and SAE [16] standards. It allows the transmitting vehicles to broadcast basic information about their position, speed, and heading. In the rest of this work, we will refer to the first message type as CAMs. The second message type represents the Maneuver Coordination Messages (MCMs) or Maneuver Sharing and Coordination Messages (MSCMs) standardized by ETSI [17] and SAE [18] to allow vehicles to share and coordinate their planned trajectories. In the rest of this work, we will refer to the second message type as MCMs. The third message type represents the Cooperative Perception Messages (CPMs) or Sensor Data Sharing Messages (SDSMs) standardized by ETSI [9] and SAE [14] for the exchange of detected objects between vehicles. In the rest of this work, we will refer to the third message type as CPMs. CPMs contain a selected subset of the exogenous variables (objects) locally detected by each transmitting vehicle through its onboard sensors[2]. The subset of locally detected exogenous variables included in each generated CPM is determined by the transmitting vehicles employing one of the following V2X communication approaches: *Baseline*, *Redundancy Mitigation (RM)*, and *Semantic*.

With the *Baseline* approach, vehicles include all locally detected exogenous variables in the generated CPM (see Fig. 2(a)). This approach is supported by both ETSI [9] and SAE [14] cooperative perception standards. With *Redundancy Mitigation*, a vehicle only transmits the locally detected exogenous variables that are not estimated to be already available (i.e., redundant) at its intended receivers (see Fig. 2(a)). A vehicle estimates the redundancy of locally detected exogenous variables by monitoring the messages exchanged on the V2X network. It assumes that any exogenous variable that it correctly receives via V2X is also available (hence, redundant) at all its intended receivers. The *Redundancy Mitigation* approach is adopted in both ETSI cooperative perception standards [9] and pre-standardization studies [19]. The *Semantic* approach follows the semantic and task-oriented V2X communication principles outlined in Section 2.1 and illustrated in Fig. 2(b). A vehicle employing the *Semantic* approach curates the content of the transmitted CPM based on its relevance for the intended receivers. An intended receiver is a vehicle that can potentially decode the transmitted CPM and is, therefore, worth to be considered when curating the message content. A vehicle $V_R$ is an intended receiver of the transmitter if the transmitter can decode at least one of the last two messages transmitted by $V_R$. A transmitting vehicle estimates the relevance of its locally detected exogenous variables for its intended receivers in two steps. First, it performs redundancy estimation (*redundancy estimation* in Fig. 1). Like in the *RM* approach, the transmitting vehicle performs redundancy estimation by monitoring the messages exchanged on the V2X network, assuming that any exogenous variable that it has correctly received is redundant for all its intended receivers. Redundant information is assumed to not improve the intended receivers' situational awareness and, therefore, redundant exogenous variables are deemed irrelevant and not included in the transmitted CPMs. Note that this is a common assumption in the literature [9][19] and is also employed by the *RM* approach. After redundancy estimation, the transmitter estimates the relevance of non-redundant locally detected exogenous variables considering all its intended receivers (*relevance estimation* in Fig. 1). As a result, the estimated relevance of each exogenous variable $x_k$ is represented as a multi-dimensional array with one estimate (relevant, irrelevant) per intended receiver. The implementation of the relevance estimation process is out of the scope of this paper and is discussed in Sec. 4.4[3]. To assess the potential scalability gains of semantic and task-oriented V2X communications, we assume that the transmitter can perfectly estimate the relevance of each locally detected exogenous variable $x_k$ for each intended receiver $V_R$. After relevance estimation, the transmitter includes in the generated CPM only those locally detected exogenous variables that are estimated to be relevant for at least one intended receiver, and discards those exogenous variables that are estimated to be irrelevant for all intended receivers (*content selection* in Fig. 1).

Vehicles exchange all generated messages using the Cellular (C)-V2X sidelink communication technology[4]. We model the probability of successfully receiving a transmitted message using the C-V2X sidelink analytical models openly released in [21]. These models provide the successful message reception probability as a function of the transmitter-receiver distance and the channel load at the receiver, taking into account the impact of packet collisions, pathloss, shadowing and fast-fading effects as well as half-duplex limitations of V2X transceivers. In C-V2X sidelink, the channel load is measured through the Channel Busy Ratio (CBR). A vehicle measures the CBR as the fraction of channel resources that it senses as occupied by other vehicles over the last 100 ms. A resource is sensed as occupied when its received signal power is larger than a pre-defined threshold. We model the sensing process using the message sensing probability analytical model also openly released in [21] and we calculate the amount of channel resources occupied by a sensed message as a function of its size and the employed Modulation and Coding Scheme (MCS). We implement the C-V2X sidelink congestion control mechanism defined by ETSI in [22], which drops (i.e., discards, does not transmit) a newly generated message if the amount of communication resources consumed by the transmitting vehicle over the last 1000 ms exceeds a pre-defined threshold that depends on the message priority and the channel load (CBR).

[2] Note that, in current V2X standards, a vehicle can include only locally collected information in the transmitted messages to avoid the propagation of errors over the V2X network.

[3] The implementation of the relevance estimation step has also been discussed in [11], while the resilience of semantic and task-oriented V2X communications to relevance estimation errors has been demonstrated in [20].

[4] Note that the operation of the proposed *Semantic* approach does not depend on the employed communication technology.

## 4. SCALABILITY ANALYSIS

We consider a 2 km x 2 km Manhattan-like driving scenario in which vehicles move at constant speed. Vehicles exchange messages on 10 MHz channels in the 5.9 GHz frequency band using the C-V2X sidelink communication technology, employing a QPSK-0.7 MCS. CAMs, MCMs, and CPMs have the same priority (i.e., PPPP5 priority according to [23]) and are periodically generated every $T$ = 100 ms, a setting which is in line with both ETSI and SAE standards. We assume that CAMs and MCMs have a fixed size and represent background traffic. The size of CAMs is set equal to 199 bytes, a value which is in line with message statistics collected by experimental studies [24] and employed in related works [11], and is representative of a CAM containing only mandatory information (e.g., transmitter's position, speed, and heading). We set the MCMs size equal to 329 bytes. This setting is representative of an MCM containing the planned trajectory of the transmitting vehicle [11]. For CPMs, the size of the message header is set equal to 121 bytes, and we assume that the size of each exogenous variable is equal to 52 bytes – the typical size of a detected object in cooperative perception [25]. The total CPM size depends on the number of locally detected exogenous variables eventually included in the generated message; hence, it depends on the employed V2X communication approach (*Baseline*, *Redundancy Mitigation*, *Semantic*). We configure each vehicle's local detection probability $P(D_{k,n})$ as follows:

$$P(D_{k,n}) = \frac{1}{1 + 0.08 \cdot e^{0.08(D_{k,n}-60)}}. \quad (1)$$

Accordingly, each vehicle's perception range is 150 m. The perception range is the distance beyond which a vehicle's probability of detecting an exogenous variable using its onboard sensors is zero. This modelling choice is in line with the object detection probability employed in other cooperative perception studies such as [5]. Unless otherwise stated, we set the relevance range $D_{max}$ equal to 400 m and the total number of relevant variables $R$ equal to 20.

### *4.1. Single-Channel Scenario*

We begin the scalability analysis focusing on a single-channel scenario where all messages (CAMs, MCMs, and CPMs) are exchanged on the same V2X channel. We assess V2X network scalability by analysing the number of vehicles that can be successfully served. A vehicle is considered successfully served if it receives, via V2X, at least $\beta$% of the relevant exogenous variables that it cannot detect with its onboard sensors but that are detected by other vehicles. Receiving such exogeneous variables is key to improve the vehicles' situational awareness beyond its onboard sensors' line-of-sight range. We define $\beta$ as a function of the distance from the receiving vehicle. To this end, the relevance range ($D_{max}$ = 400 m) is divided into four relevance zones, each reflecting a different level of urgency. Zone 1 covers distances up to 100 m and includes relevant exogenous variables with critical urgency that may represent an immediate safety threat to the receiving vehicle (e.g., an imminently colliding object). Accordingly, we set $\beta$ = 99% in Zone 1 – a configuration that is in line with the typical service-level requirements defined by ETSI [26] and 5GAA [27] for safety-critical use cases. Zone 2 (high urgency), Zone 3 (medium urgency), and Zone 4 (low urgency) respectively cover the 100-200 m, 200-300 m, and 300-400 m distance ranges and represent zones with increasingly less urgent information. Accordingly, their $\beta$ requirements are progressively relaxed. We respectively set $\beta$ equal to 75%, 60%, and 50% in zones 2, 3, and 4. This distance-dependent definition of $\beta$ provides a simple yet effective way to model the varying levels of service required by a connected vehicle.

Fig. 3 reports the probability *P(succ)* that a vehicle is successfully served in Zone 1 as a function of the number of communicating vehicles, $N$, considering three different exogenous variables densities. The three densities correspond to an average number of locally detected exogenous variables, $M$, equal to 10, 20, and 30. In the cooperative perception context, $M$ = 10 represents a scenario with a low object density that corresponds, for example, to a three-lane-per-direction highway with low traffic density (e.g., 20 veh/km) [28]. $M$ = 20 corresponds to a medium object density scenario, representative of a five-lane-per-direction highway with low traffic density [28], for example. Finally, $M$ = 30 represents a high object density scenario, as found in suburban areas involving both vehicles and vulnerable road users (e.g., pedestrians and cyclists) [29], for example.

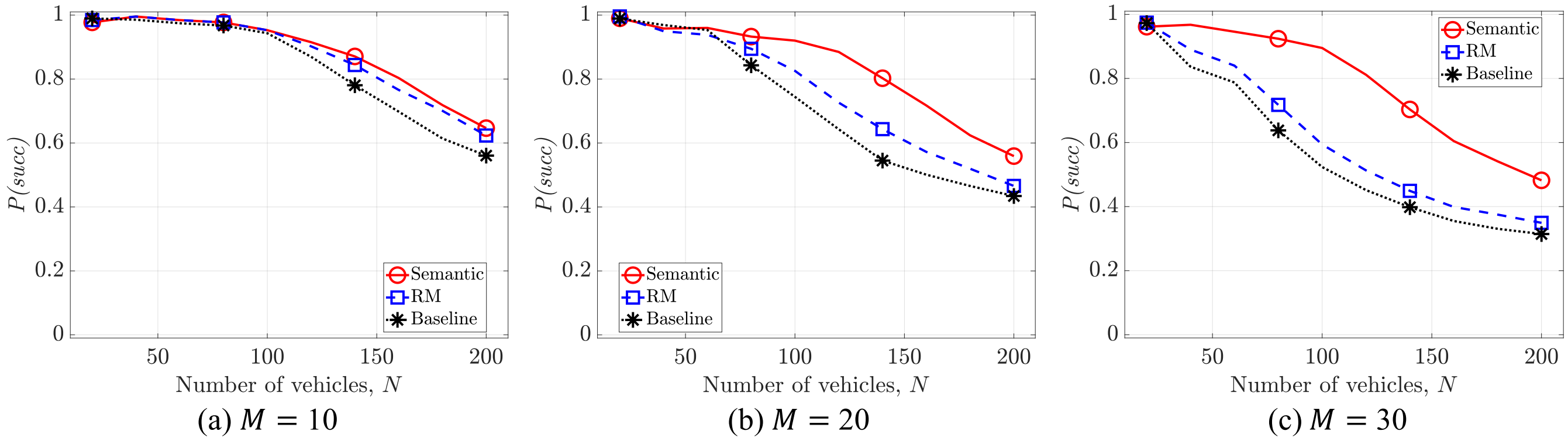

(a) $M = 10$ (b) $M = 20$ (c) $M = 30$

**Fig. 3**. Single-channel scenario, $M = 10$, 20 and 30. Probability *P(succ)* that a vehicle is successfully served. Zone 1 (0-100 m, $\beta = 99\%$).

Fig. 3 shows (i) that semantic and task-oriented V2X communications can improve the probability *P(succ)* that a vehicle is successfully served, and (ii) that the improvement depends on the exogenous variables density. When $M = 10$ (Fig. 3(a)), the *Semantic* approach does not significantly improve *P(succ)* with respect to its non-semantic (*RM* and *Baseline*) counterparts. This is the case because the number of exogenous variables locally detected by each vehicle is small and a relevance-aware content selection has limited impact. This impact becomes more evident as $M$ increases, in Fig. 3(b) and Fig. 3(c). When $M = 30$ (Fig. 3(c)), the *Semantic* approach is able to guarantee *P(succ)* values that are up to 75% larger with respect to the *Redundancy Mitigation* and *Baseline* approaches. Higher *P(succ)* values indicate that vehicles have higher probability of successfully receiving the relevant information that is critical to their driving (Zone 1). To quantify the scalability gains achieved with semantic and task-oriented V2X communications, we derive from Fig. 3 the number of successfully served vehicles with a 90% probability (i.e., with *P(succ)* = 0.9). The results show that, in Zone 1, the *Semantic* approach can increase by a factor of 2.8 and 3.0 the number of successfully served vehicles compared to its *Redundancy Mitigation* and *Baseline* counterparts. We select the 90% probability as a sufficiently high threshold to consider the communicating vehicles as reliably served with the required relevant information.

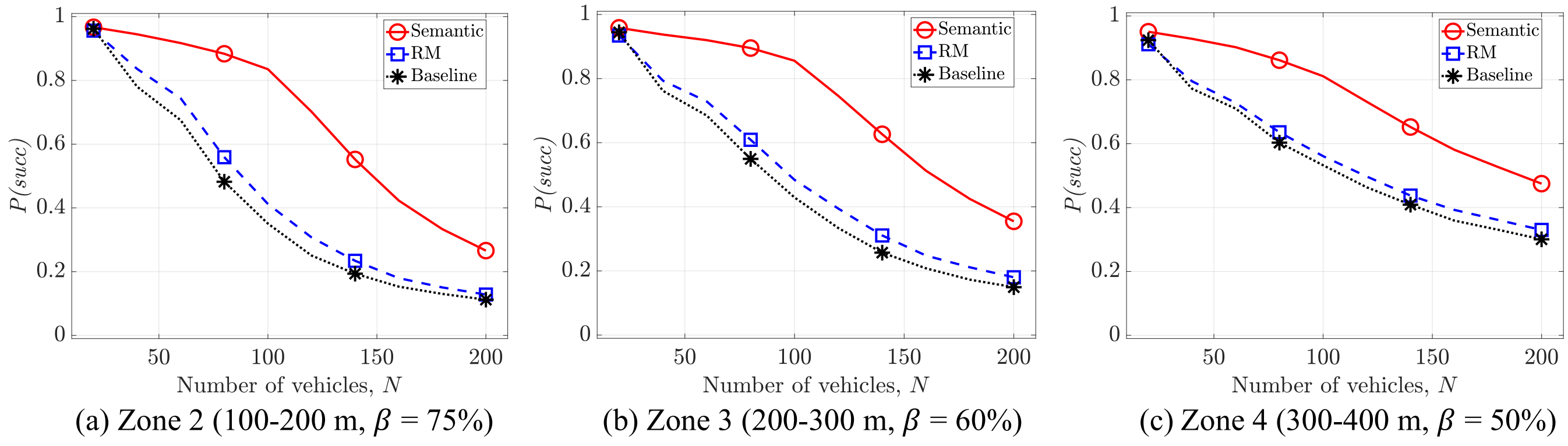

(a) Zone 2 (100-200 m, $\beta = 75\%$) (b) Zone 3 (200-300 m, $\beta = 60\%$) (c) Zone 4 (300-400 m, $\beta = 50\%$)

**Fig. 4**. Single-channel scenario, $M = 30$. Probability *P(succ)* that a vehicle is successfully served. Zones 2, 3, and 4.

Fig. 4 completes the scalability analysis in the high-density scenario ($M = 30$) focusing on Zone 2 (Fig. 4(a)), Zone 3 (Fig. 4(b)), and Zone 4 (Fig. 4(c)). Fig. 4 demonstrates that semantic and task-oriented V2X communications consistently guarantee higher *P(succ)* values to the receiving vehicles across all zones. Moreover, this figure shows that the *P(succ)* gap between *Semantic* and non-semantic (*RM* and *Baseline*) approaches decreases in relevance zones farther from the receiving vehicle. As the distance increases, the impact of propagation losses on the correct reception of relevant information becomes increasingly important and cannot be mitigated by the reduction of packet collisions achieved with more effective V2X communication approaches (*Semantic*) that decrease the channel load. Nevertheless, semantic and task-oriented V2X communications are able to provide significant scalability gains compared to existing communication approaches in all zones. In particular, the *Semantic* approach can successfully serve – with a 90% probability – 2.7, 2.6, and 2.3 times more vehicles in Zone 2, 3, and 4, respectively, compared to non-semantic approaches.

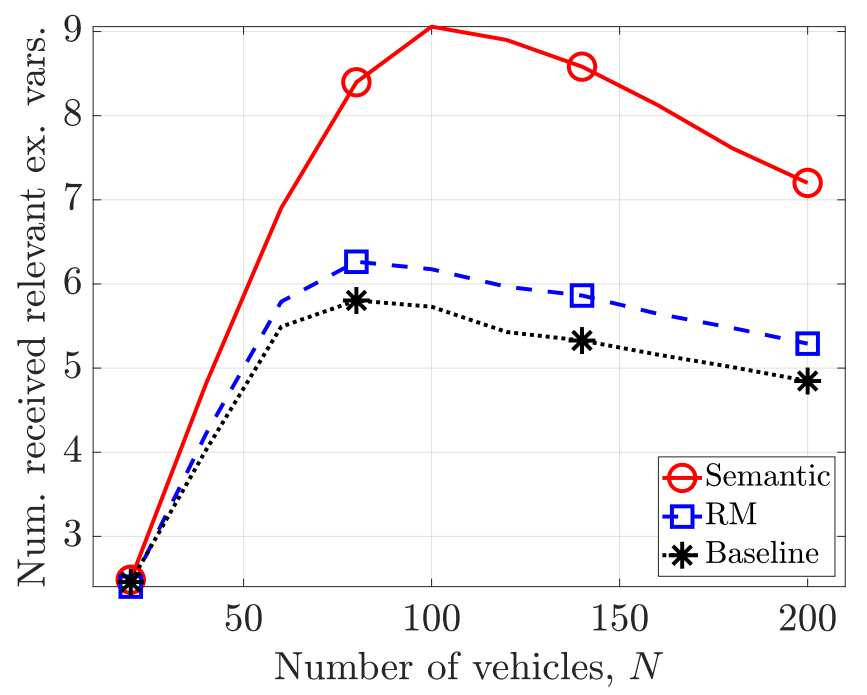

(a) Number of received relevant ex. vars.

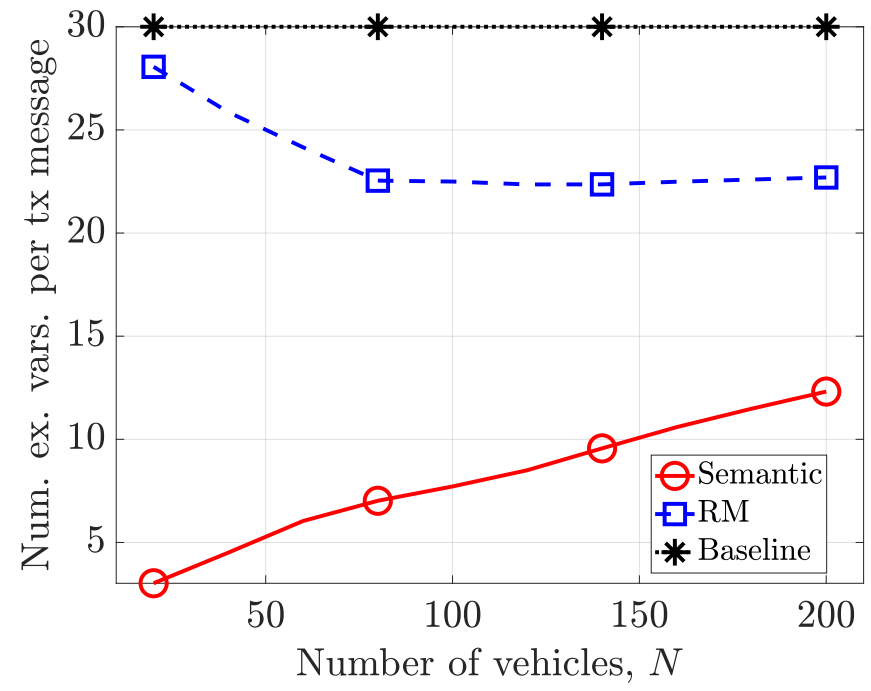

(b) Number of ex. vars. per transmitted message

**Fig. 5**. Single-channel scenario, $M = 30$. Average number of received relevant exogenous variables and average number of exogenous variables per transmitted message.

The scalability gains achieved with the *Semantic* approach across all relevance zones (see Fig. 3 and Fig. 4) stem from the enhanced communication efficiency of semantic and task-oriented V2X communications. With semantic and task-oriented V2X communications, vehicles can deliver larger amounts of relevant information to their intended receivers while consuming less communication resources. This is visible in Fig. 5(a) and Fig. 5(b), which report the average number of relevant exogenous variables received by each vehicle via V2X and the average number of exogenous variables per transmitted message – a proxy of the CPMs size – measured in the high-density scenario ($M = 30$). Fig. 5(a) clearly shows that the *Semantic* approach guarantees the delivery of larger amounts of relevant information with respect to non-semantic approaches (*RM* and *Baseline*), regardless of the number of communicating vehicles $N$. Fig. 5(b) demonstrates that this is actually achieved while minimizing the number of exogenous variables per transmitted message. By focusing on the transmission of relevant information, the *Semantic* approach can prevent the transmission of information that is not needed at the intended receivers and, as a result, greatly improve the communication efficiency.

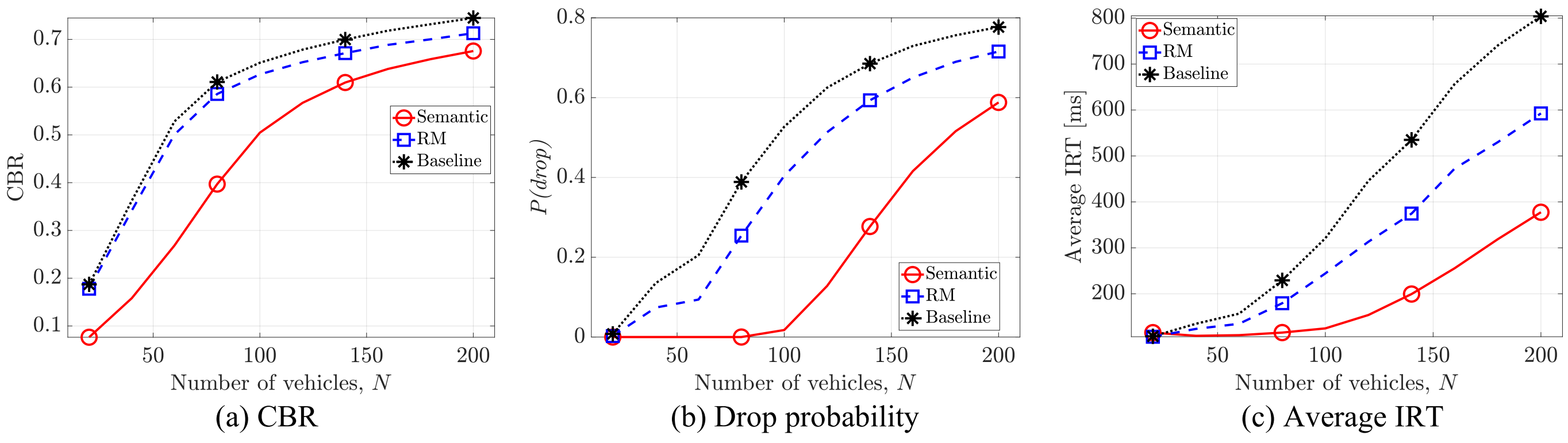

(a) CBR (b) Drop probability (c) Average IRT

**Fig. 6**. Single-channel scenario, $M = 30$. CBR, drop probability, and average IRT.

The message size reduction enabled by semantic and task-oriented V2X communications directly translates into a lower CBR (i.e., network load)[5]. This is illustrated in Fig. 6(a) for the $M = 30$ scenario. A lower CBR reduces both the message collision probability and the message drop probability, *P(drop)*, as shown by Fig. 6(b). Here, *P(drop)* denotes the probability that a generated message is dropped by the congestion control mechanism at the transmitting vehicle to limit the CBR. Fig. 6(b) shows that, in the *Redundancy Mitigation* and *Baseline* case, *P(drop)* is larger than zero even at very low vehicular densities ($N < 50$). In contrast, *P(drop)* becomes larger than zero only when $N > 100$ with the *Semantic* approach. Note that such *P(drop)* increase is responsible for the reduction in the number of received relevant exogenous variables reported in Fig. 5(a) for $N > 100$.

A lower drop probability indicates that each vehicle can transmit its generated messages more frequently and guarantee more frequent updates at the intended receivers. This effect is visible in Fig. 6(c), which analyzes the average message Inter-Reception Time (IRT). At the receiver, the IRT is defined as the time interval between the successful reception of two consecutive messages from the same transmitting vehicle. Fig. 6(c) shows that, on average, a vehicle may wait up to 800 ms or 600 ms to receive a new message from a transmitting vehicle using the *Baseline* and *Redundancy Mitigation* approaches, respectively. In contrast, the *Semantic* approach guarantees a maximum IRT below 400 ms – a 50% and 33% reduction compared to the *Baseline* and *Redundancy Mitigation* approaches. This improvement is particularly significant given that vehicles generate a new message every $T = 100$ ms. Operating with a low IRT is especially important in safety-critical V2X applications that require frequent and regular updates. Overall, the results in Fig. 6(c) highlight the two-fold benefits of semantic and task-oriented V2X communications: they improve V2X network scalability by successfully serving a larger number of vehicles and, at the same time, guarantee a lower IRT to the communicating vehicles.

[5] It is worth highlighting that the vehicular densities and the CBR values considered in this study are in line with those of other existing works such as [30].

### *4.2. Single-Channel Scenario – Sensitivity Analysis*

This section analyses the sensitivity of the scalability gains achieved with semantic and task-oriented V2X communications to different driving conditions. We represent different driving conditions by varying two main parameters: $R$ and $D_{max}$. Recall $D_{max}$ represents the maximum distance between a vehicle and a relevant exogenous variable while $R$ denotes the total number of relevant exogenous variables for each vehicle.

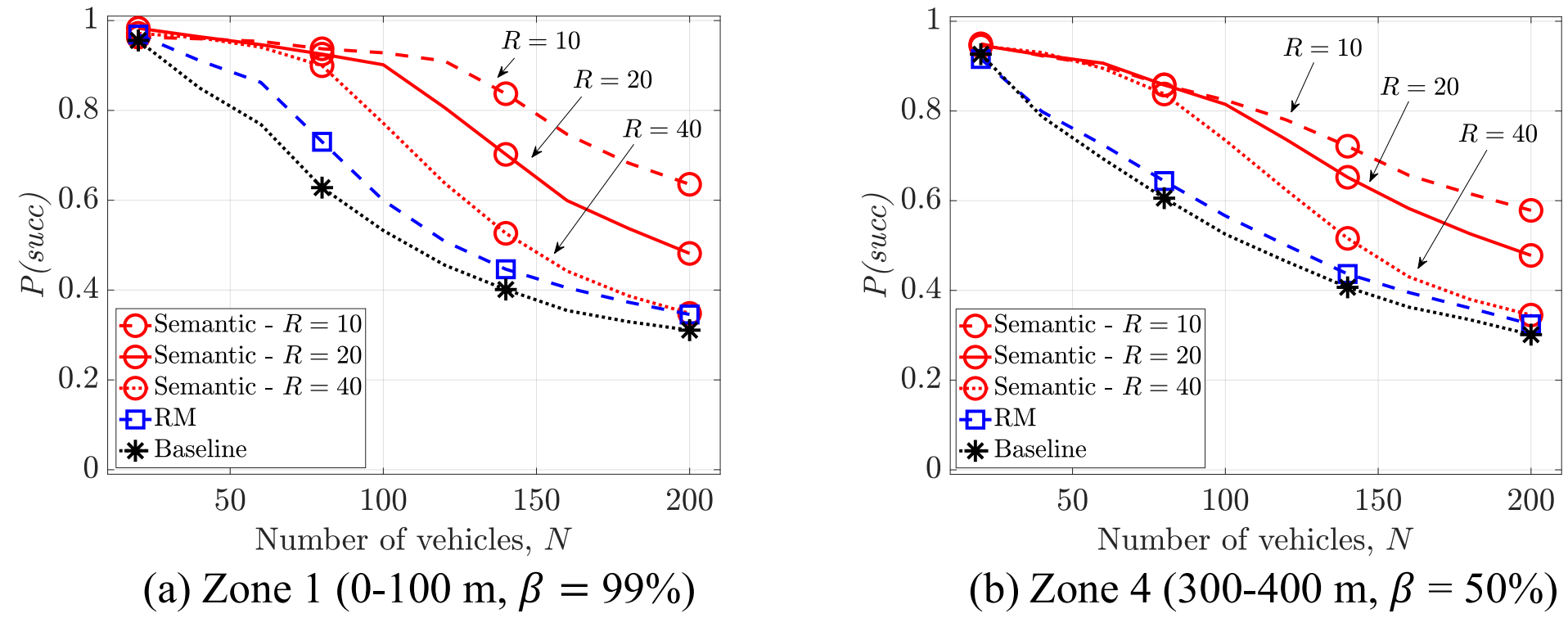

(a) Zone 1 (0-100 m, $\beta = 99\%$) (b) Zone 4 (300-400 m, $\beta = 50\%$)

**Fig. 7**. Single-channel scenario, $M = 30$ and varying $R$. Probability *P(succ)* that a vehicle is successfully served. Zones 1 and 4.

Variations in $R$ capture the complexity of different driving scenarios. Low values of $R$ correspond to low-complexity driving scenarios where only a few objects (exogenous variables) can influence a vehicle's driving decisions. This is the case, for example, of a free-flow highway. In contrast, high values of $R$ represent high-complexity scenarios, such as dense urban environments, where a vehicle must pay attention to multiple objects, including pedestrians, bicycles and other vulnerable road users in addition to other surrounding vehicles. Fig. 7 analyses the impact of $R$ on *P(succ)*, the probability that a vehicle is successfully served, focusing on Zone 1 and Zone 4. Similar trends are observed in other zones (Zone 2 and 3). Fig. 7 shows that smaller values of $R$ widen the performance gap between the *Semantic* approach and non-semantic approaches (*RM* and *Baseline*) in both Zone 1 (Fig. 7(a)) and Zone 4 (Fig. 7(b)). In Zone 1, the *Semantic* approach achieves up to a two-fold increase in *P(succ)* compared to the *Redundancy Mitigation* and *Baseline* approaches. Notably, $R$ does not affect the *Redundancy Mitigation* and *Baseline* performance, as these approaches do not account for the relevance of the exchanged information when curating the transmitted messages content. The improved *P(succ)* performance indicates a higher probability of successfully delivering the required relevant information to the intended receivers and, therefore, greater scalability gains of semantic and task-oriented V2X communications as smaller values of $R$ are considered. In Zone 1, the number of vehicles that can be successfully served by *Semantic* is 2.5, 3.0, and 4.1 times larger with respect to the *Baseline* approach when $R$ = 40, 20, and 10, respectively.

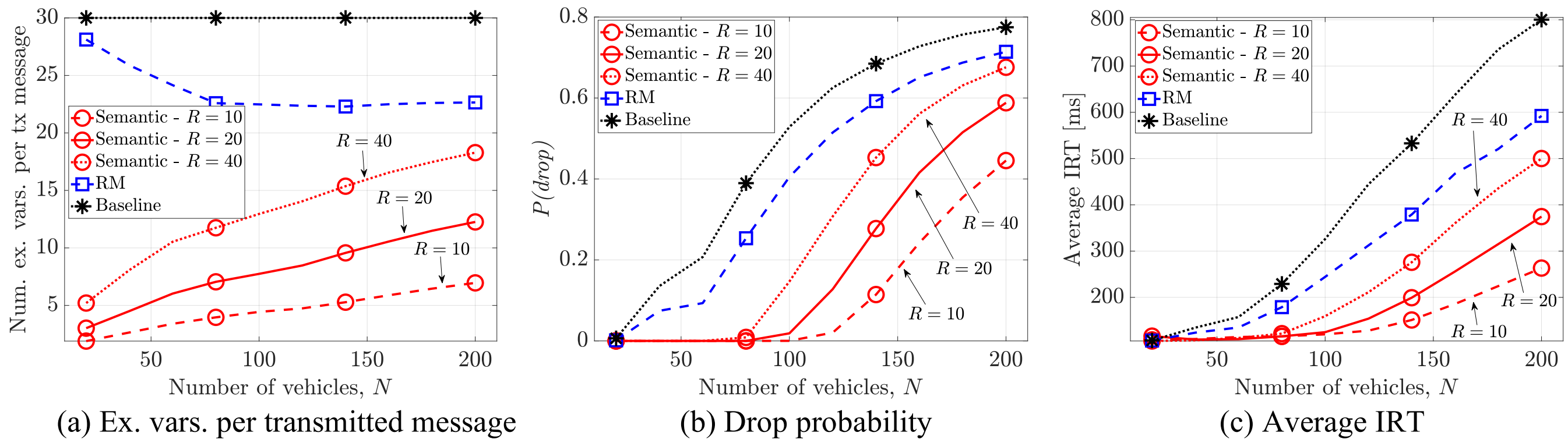

(a) Ex. vars. per transmitted message (b) Drop probability (c) Average IRT

**Fig. 8**. Single-channel scenario, $M = 30$ and varying $R$. Average number of exogeneous variables per transmitted message, drop probability, and average IRT.

Varying $R$ primarily affects the number of exogenous variables (objects) that each vehicle must transmit to deliver the required relevant information to its intended receivers. This is illustrated in Fig. 8(a), where the *Semantic* approach reduces the number of exogenous variables per message as $R$ decreases, while the *Redundancy Mitigation* and *Baseline* don't. As already highlighted by Fig. 6, reducing the number of exogenous variables per transmitted message lowers the message drop probability and, consequently, the average IRT, as respectively shown by Fig. 8(b) and Fig. 8(c). As a result, the average IRT of the *Semantic* approach is confined below 263 ms in the $R = 10$ case, a 67% and 56% reduction with respect to the average IRT levels achieved with the *Redundancy Mitigation* and *Baseline* approaches. Varying values of $R$ simultaneously affect the scalability gains of semantic and task-oriented V2X communications as well as the time between the reception of consecutive updates.

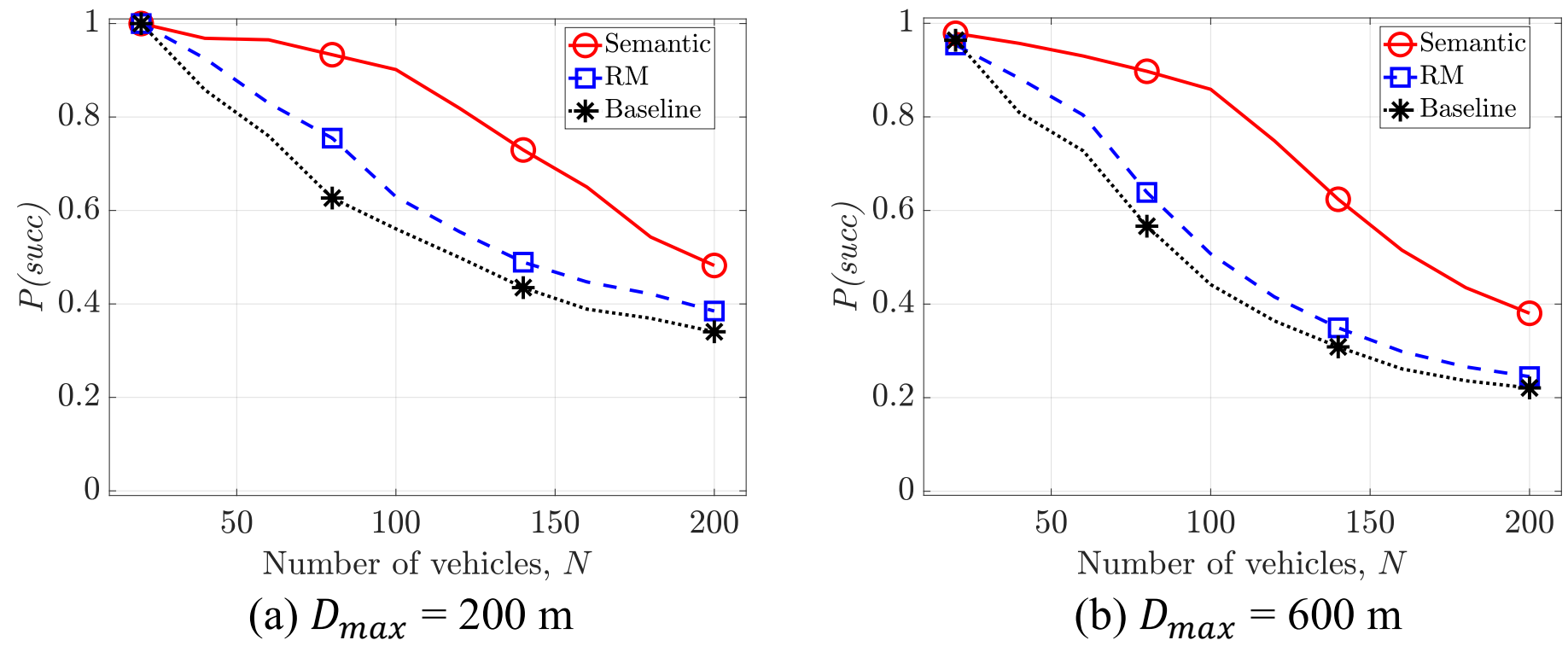

(a) $D_{max} = 200$ m (b) $D_{max} = 600$ m

**Fig. 9**. Single-channel scenario, $M = 30$ and varying $D_{max}$. Probability *P(succ)* that a vehicle is successfully served. Zone 1.

Variations in $D_{max}$ capture the impact of the vehicle's speed on the maximum distance between the vehicle and relevant exogenous variables (objects). Larger values of $D_{max}$ correspond to high-speed conditions (e.g., 120 km/h), where a vehicle's driving may potentially be influenced by distant objects. In contrast, smaller $D_{max}$ values represent low-speed driving conditions (e.g., 40 km/h) where a vehicle's driving can only be influenced by surrounding objects in its close proximity. Fig. 9 reports the probability *P(succ)* that a vehicle is successfully served in Zone 1 for two different values of $D_{max}$, 200 m (Fig. 9(a)) and 600 m (Fig. 9(b)). Note that variations in $D_{max}$ accordingly modify the range of distances including critical-urgency information and covered by Zone 1. Zone 1 covers distances up to 50 m when $D_{max}$ = 200 m and up to 150 m when $D_{max}$ = 600 m. Fig. 9 shows that variations in $D_{max}$ do not substantially affect the *P(succ)* performance gap between *Semantic* and non-semantic (*RM* and *Baseline*) approaches. Semantic and task-oriented V2X communications consistently achieve a more effective delivery of relevant information in Zone 1, maintaining their scalability gains in the most critical zone across different $D_{max}$ settings. The *Semantic* approach achieves an approximately 3.0x increase in the number of successfully served vehicles compared to its *Redundancy Mitigation* and *Baseline* counterparts in both Fig. 9(a) and Fig. 9(b). Note that these scalability gains are in line with those reported in Fig. 3(c) where $D_{max}$ = 400 m.

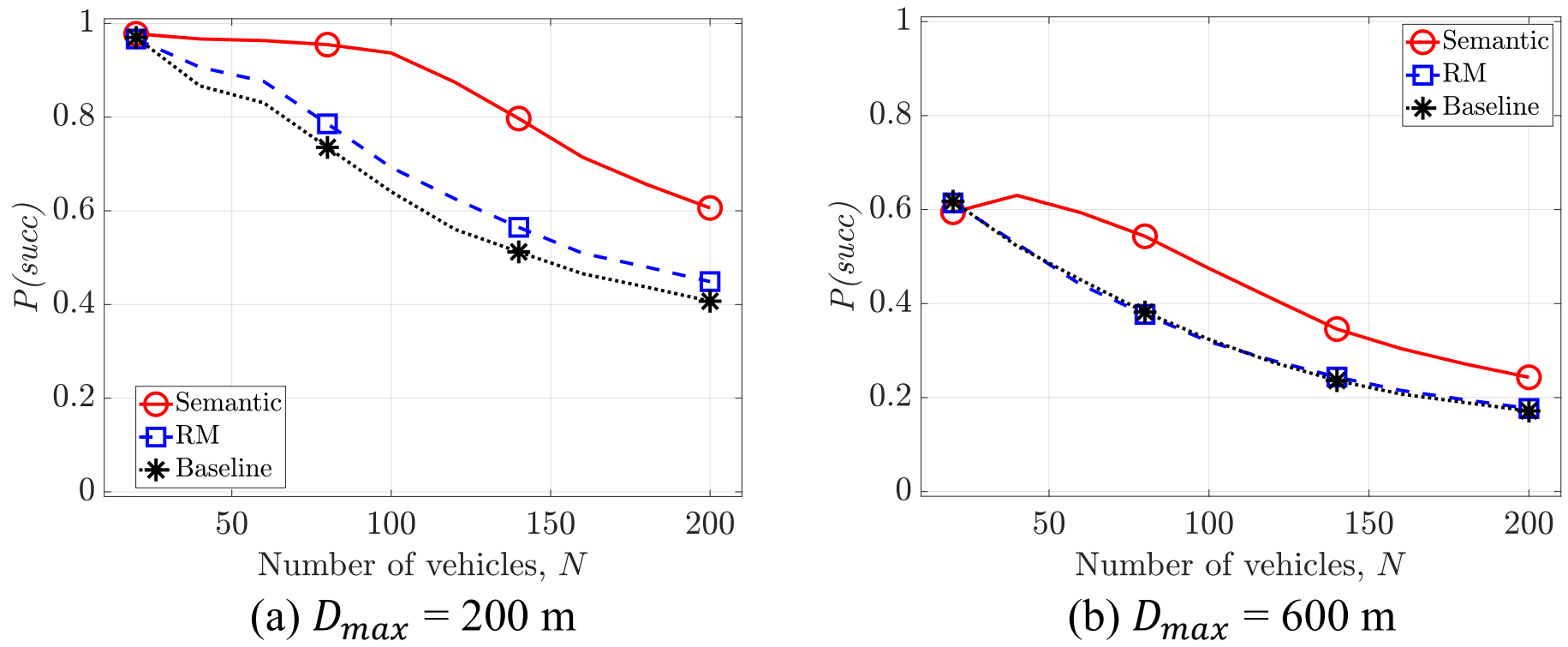

(a) $D_{max}$ = 200 m (b) $D_{max}$ = 600 m

**Fig. 10**. Single-channel scenario, $M = 30$ and varying $D_{max}$. Probability *P(succ)* that a vehicle is successfully served. Zone 4.

It is worth highlighting that, while larger $D_{max}$ values do not affect the scalability gains of semantic and task-oriented V2X communications in Zone 1, they can influence the successful delivery of relevant information in farther – and less critical – relevance zones. This is because larger $D_{max}$ values increase the distance between the receiving vehicle and the vehicles reporting relevant information, thereby increasing the impact of propagation losses on the correct delivery of relevant information. In Zone 1, where the range of distances closer to the receiving vehicle are considered, this impact is negligible. In Zone 4, it becomes more evident as highlighted by Fig. 10, where the *P(succ)* performance in Zone 4 is analysed in both the $D_{max}$ = 200 m and $D_{max}$ = 600 m settings. Note that Zone 4 covers the 150-200 m range of distances when $D_{max}$ = 200 m and the 450-600 m interval when $D_{max}$ = 600 m. The comparison between Fig. 10(a) and Fig. 10(b) clearly shows that, in Zone 4, *P(succ)* gets strongly deteriorated at larger $D_{max}$ values and remains below 0.6 even with the *Semantic* approach, indicating that a more efficient usage of communication resources is not sufficient to mitigate the impact of propagation losses.

### *4.3. Multi-Channel Scenario*

This section investigates the scalability gains of semantic and task-oriented V2X communications considering a multi-channel scenario. In the multi-channel scenario, we assume that the V2X network features three different V2X channels and that each type of generated message (CAM, MCM, or CPM) is exchanged on a dedicated channel.

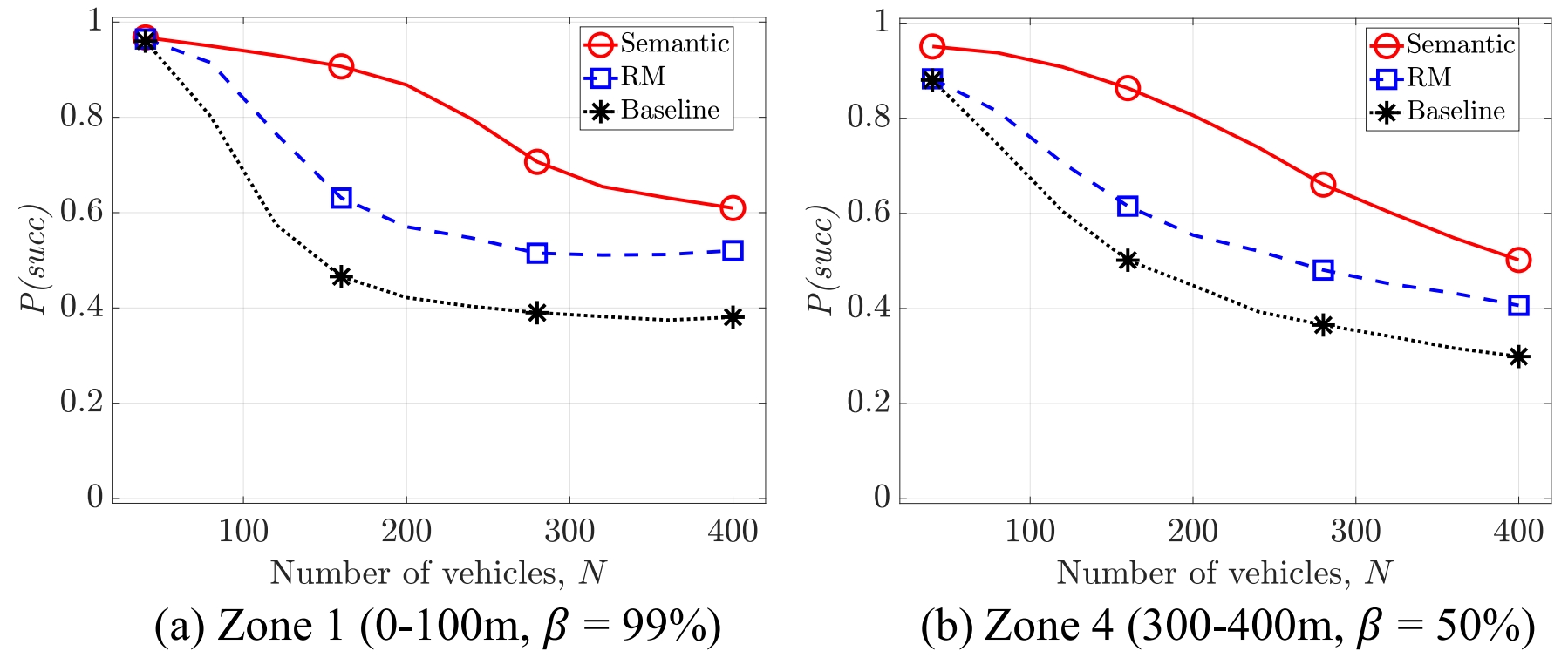

(a) Zone 1 (0-100m, $\beta$ = 99%) (b) Zone 4 (300-400m, $\beta$ = 50%)

**Fig. 11**. Multi-channel scenario, $M = 30$. Probability *P(succ)* that a vehicle is successfully served. Zones 1 and 4.

Focusing on the high-density scenario ($M$ = 30), Fig. 11 reports the probability *P(succ)* that a vehicle is successfully served in the most (Zone 1) and least (Zone 4) critical relevance zones, as a function of $N$. The results in Fig. 11 show that the *Semantic* approach consistently achieves higher *P(succ)* values than its non-semantic (*RM* and *Baseline*) counterparts across both relevance zones and for all considered vehicular densities[6]. This indicates that semantic and task-oriented V2X communications can consistently enhance the V2X network scalability across different relevance zones also in the multi-channel scenario. In particular, the per-zone comparison between the multi-channel and single-channel scenarios reveals that the scalability gains of semantic and task-oriented V2X communications are largely maintained – actually, slightly increased – in the multi-channel case. In the single-channel scenario, the *Semantic* approach successfully serves – with a 90% probability – a number of vehicles that is 3.0 times higher with respect to the *Baseline* approach in Zone 1 (Fig. 3(c)). This gain increases to a factor of 3.2 in the multi-channel scenario (Fig. 11(a)). Similar trends are observed in Zone 4 and also in other relevance zones (Zone 2 and Zone 3). In Zone 2, the scalability gain increases from a 2.7x factor in the single-channel scenario to 2.9x in the multi-channel scenario. In Zone 3, it grows from 2.6x (single-channel) to 2.9x (multi-channel), while in Zone 4 it increases from 2.3x (single-channel, Fig. 4(c)) to 2.6x (multi-channel, Fig. 11(b)). The larger gains in the multi-channel scenario stem from the absence of background traffic, which allows semantic and task-oriented V2X communications to affect a larger fraction of the traffic exchanged on the channel. In contrast, the scalability gains of the *Semantic* approach over its *Redundancy Mitigation* counterpart do not modify in the multi-channel scenario. In both the single-channel and multi-channel scenarios, the *Semantic* approach successfully serves 2.8x, 2.7x, 2.6x, and 2.3x more vehicles than the *Redundancy Mitigation* approach in Zones 1, 2, 3, and 4, respectively.

[6] Note that the vehicular densities ($N$) analysed in the multi-channel scenario are larger with respect to the single-channel scenario. This is because, in the multi-channel scenario, CPMs are exchanged over a dedicated channel with no background traffic and, therefore, a larger number of vehicles is required to saturate the V2X channel.

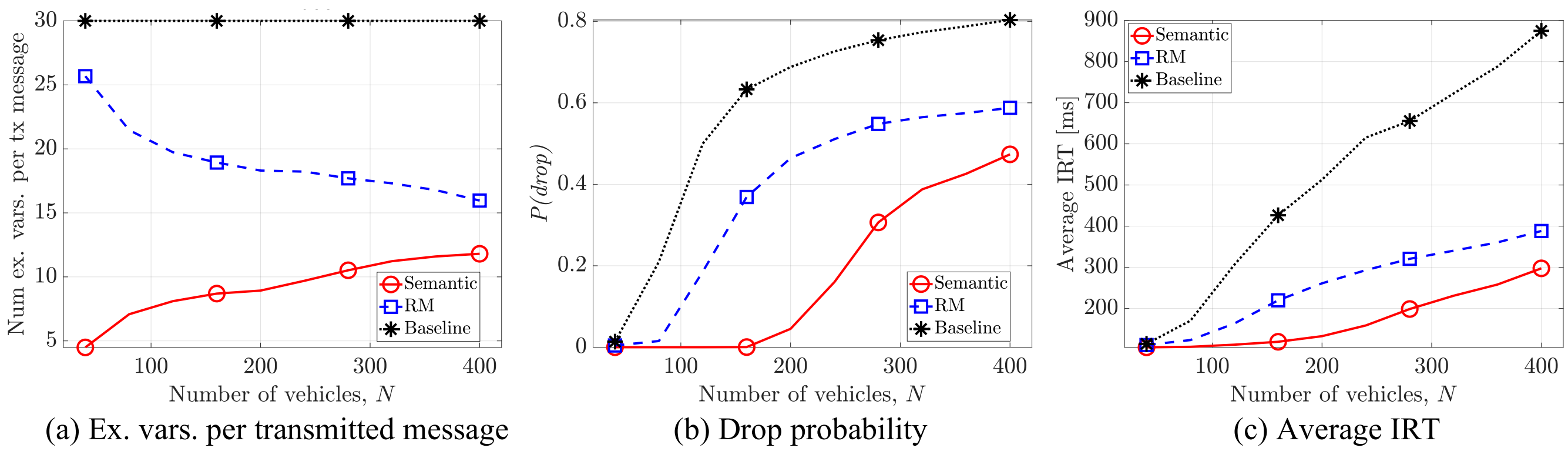

(a) Ex. vars. per transmitted message (b) Drop probability (c) Average IRT

**Fig. 12**. Multi-channel scenario, $M = 30$. Average number of exogeneous variables per transmitted message, drop probability, and average IRT.

In the multi-channel scenario, the *Redundancy Mitigation* approach greatly decreases the number of exogenous variables per transmitted message (Fig. 12(a)) with respect to the single-channel scenario (Fig. 5(b)) given the higher vehicular densities and the resulting increase in redundancy; yet, it still does not match the performance of the *Semantic* approach. The comparison between the *Semantic* and *Redundancy Mitigation* approaches in Fig. 12(a) further confirms that filtering the content of the transmitted messages based solely on redundancy is not sufficient to prevent the transmission of unnecessary information and to minimize the consumption of communication resources. By focusing on relevance, the *Semantic* approach minimizes the number of exogenous variables per transmitted message – hence, the communication resources usage – also in the multi-channel scenario. As already highlighted in the single-channel scenario, a more effective relevance-based *Semantic* V2X communication approach not only improves V2X network scalability, but it also has a positive effect on the drop probability and, ultimately, on the average IRT. The same trend is maintained for the multi-channel scenario. Fig. 12(b) reports the drop probability as a function of $N$ in the multi-channel scenario, showing that the *Semantic* approach is able to greatly decrease the drop probability with respect to non-semantic approaches, especially at low-medium vehicular densities ($N < 300$). Fig. 12(c) shows that the *Semantic* approach is able to decrease the average IRT also in the multi-channel scenario and guarantee a more regular and frequent delivery of information at the intended receivers. The IRT attained by the *Baseline* and *Redundancy Mitigation* approaches can be as high as 874 ms and 388 ms, respectively. In contrast, the *Semantic* approach confines the average IRT below 297 ms – a 66% and 22% reduction with respect to its non-semantic counterparts.

### *4.4. Discussion*

Our analysis reveals that, by focusing on relevance, semantic and task-oriented V2X communications can significantly enhance the communication efficiency and, consequently, improve the scalability of V2X networks with respect to existing V2X communication approaches. The effectiveness of semantic and task-oriented V2X communications depends on the transmitting vehicles' capability to correctly identify and transmit only the information that is relevant to their intended receivers. However, relevance estimation can be a particularly challenging task in practical V2X deployment scenarios. Driving environments are highly dynamic and involve complex interactions among vehicles, as well as between vehicles and other road users. These conditions can hinder an accurate estimation of the intended receivers' context that, in turn, can directly lead to relevance estimation errors.

Yet, the V2X domain represents a unique ecosystem that inherently provides all necessary tools to address practical relevance estimation challenges and realize the potential of semantic and task-oriented V2X communications. First, the V2X domain is rich of contextual information about both the driving environment and the communicating vehicles. Vehicles continuously sense the surrounding environment and exchange sensing data, as well as information about their current state (e.g., position and speed) and future intentions (planned trajectories). This information is often enriched by the roadside infrastructure, which can provide additional or higher quality data. Second, connected vehicles are powerful mobile computing platforms equipped with the hardware and software capabilities required to support the real-time execution of sophisticated AI-based relevance estimation algorithms. These algorithms are key to process and interpret the available contextual information and ultimately allow an accurate relevance estimation. Third, the broadcast nature of V2X networks offers an inherent resilience to potential relevance estimation errors. As demonstrated in [20], if a vehicle fails to transmit a relevant piece of information due to a relevance estimation error, another nearby vehicle can naturally compensate by delivering the missing information to the intended receivers.

## 5. CONCLUSIONS

Semantic and task-oriented V2X communications can improve the communication efficiency by focusing on the transmission of relevant information. In this study, we analysed the potential V2X network scalability gains offered by semantic and task-oriented V2X communications under different driving conditions and in both single and multi-channel V2X scenarios. The obtained results numerically demonstrate that semantic and task-oriented V2X communications can substantially improve the V2X networks scalability compared to existing V2X communication approaches, particularly when the delivery of information with a critical urgency for the intended receivers is considered. In particular, results show that semantic and task-oriented V2X communications can increase by up to a 4.1x factor the number of successfully served vehicles in high-density single-channel scenarios and maintain these scalability gains in multi-channel scenarios, which are currently considered for future V2X deployment phases. Furthermore, our analysis reveals that semantic and task-oriented V2X communications also decrease the inter-reception time between consecutive messages by up to 67%, ensuring a more regular and frequent delivery of relevant information at the intended receivers – a critical requirement for V2X-enabled safety applications.